\documentclass[aip,apl,amsmath,amssymb,reprint,graphicx]{revtex4-1}
\usepackage{graphicx}
\usepackage{subfigure}
\usepackage{siunitx}
\usepackage{physics}
\usepackage[normalem]{ulem} 

\usepackage[usenames,dvipsnames]{color} 

\definecolor{orange}{rgb}{0.7,0.2,0}

\definecolor{darkgreen}{rgb}{0,0.3,0}

\newcommand{\droh}[1]{\dot{\rho}_{#1}}
\newcommand{\ima}{\mathrm{i}}
\newcommand{\lnv}{\Lambda_{\mathrm{NV}}}
\newcommand{\gnv}{G_{\mathrm{NV}}}
\newcommand{\lij}[1]{L_{#1}}
\draft 

\begin{document}


\title[Two-media laser threshold magnetometry: A magnetic-field-dependent laser threshold]{Two-media laser threshold magnetometry: A magnetic-field-dependent laser threshold} 



\author{Yves Rottstaedt$^\dagger$}
\email[]{yves.rottstaedt@uni-leipzig.de}
\thanks{$^\dagger$These authors contributed equally to this work.}
\affiliation{Fraunhofer-Institut für Angewandte Festkörperphysik (IAF), Tullastraße 72, 79108
Freiburg, Germany}
\affiliation{Institut für Theoretische Physik, Universität Leipzig, Brüderstraße 16, 04103 Leipzig, Germany}

\author{Lukas Lindner$^\dagger$}
\affiliation{Fraunhofer-Institut für Angewandte Festkörperphysik (IAF), Tullastraße 72, 79108
Freiburg, Germany}

\author{Florian Schall}
\affiliation{Fraunhofer-Institut für Angewandte Festkörperphysik (IAF), Tullastraße 72, 79108
Freiburg, Germany}

\author{Felix A. Hahl}
\affiliation{Fraunhofer-Institut für Angewandte Festkörperphysik (IAF), Tullastraße 72, 79108
Freiburg, Germany}

\author{Tingpeng Luo}
\affiliation{Fraunhofer-Institut für Angewandte Festkörperphysik (IAF), Tullastraße 72, 79108
Freiburg, Germany}

\author{Florentin Reiter}
\affiliation{Fraunhofer-Institut für Angewandte Festkörperphysik (IAF), Tullastraße 72, 79108
Freiburg, Germany}

\author{Takeshi Ohshima}
\affiliation{National Institutes for Quantum Science and Technology
(QST), 1233 Watanuki, Takasaki, Gunma 370-1292, Japan}
\affiliation{Department of Materials Science, Tohoku University, Aoba, Sendai, Miyagi, 980-8579, Japan}

\author{Alexander M. Zaitsev}
\affiliation{College of Staten Island,
CUNY, 2800 Victory Blvd., Staten Island, NY 10312, USA}

\author{Roman Bek}
\affiliation{Twenty-One Semiconductors GmbH, Kiefernweg 4, 72654 Neckartenzlingen, Germany}

\author{Marcel Rattunde}
\affiliation{Fraunhofer-Institut für Angewandte Festkörperphysik (IAF), Tullastraße 72, 79108
Freiburg, Germany}

\author{Jan Jeske}
\email[]{jan.jeske@iaf.fraunhofer.de}
\affiliation{Fraunhofer-Institut für Angewandte Festkörperphysik (IAF), Tullastraße 72, 79108
Freiburg, Germany}

\author{Rüdiger Quay}
\affiliation{Fraunhofer-Institut für Angewandte Festkörperphysik (IAF), Tullastraße 72, 79108
Freiburg, Germany}


\date{\today}
\begin{abstract}
Nitrogen-vacancy (NV) centers in diamond are a promising platform for high-precision magnetometry. 
In contrast to the use of spontaneous emission in a number of NV-magnetometers, laser threshold magnetometry 
(LTM) exploits stimulated emission of NV centers by placing an NV-doped diamond inside an optical cavity. 
The NV laser system is predicted to reach a high magnetic-field-dependent contrast and coherent signal strength, leading 
to an improved magnetic field sensitivity combined with a high linearity. Here, we consider a two-media 
setup where the cavity additionally includes a vertical external cavity surface emitting laser. This optically active 
material compensates cavity losses at \SI{750}{nm} while still allowing for magnetic-field-dependent effects from the NV-diamond. 
We demonstrate a magnetic-field-dependent laser threshold and investigate the effects of pump laser induced absorption of the diamond. 
The experimental data is supported by an analytical simulation based on a rate model. Furthermore, we derive a generalized 
formula to compute the shot-noise-limited magnetic field sensitivity in the regime of high contrast yielding 
$\SI{49.07 \pm 0.33}{\pico\tesla\per\sqrt{Hz}}$ for the present setup. Simulations with an optimized NV-diamond suggest 
that values down to \SI{4.9}{\femto\tesla\per\sqrt{Hz}} are possible.

\end{abstract}


\maketitle 

\section{Introduction}
Measuring magnetic fields has become increasingly important as quantum sensors enable applications in multiple fields of 
research~\cite{Degen2017}. The improved precision allows, for example, the measurement of biomagnetic signals from the brain or the 
heart~\cite{Boto2018,Zhang2020,Bison2009,Lembke2014}, the study of magnetic structures and 
anomalies~\cite{Keenan2011}, and fundamental research of magnetism~\cite{McCullian2020}. \\
The predominant methods for high-precision magnetometry are superconducting quantum interference devices (SQUIDs) and optically 
pumped vapor cell magnetometers (OPMs). While it is possible to reach outstanding sensitivities ($<\SI{1}{\femto\tesla\per\sqrt{Hz}}$~\cite{Dang2010,Simmonds1979}) with these, 
SQUIDs require cryogenic temperatures, which leads to setups with a large periphery and OPMs need to be heated~\cite{Degen2017,Aslam2023}. \\
As an alternative, nitrogen-vacancy (NV) centers in diamond have been established~\cite{Xu2023}. The NV center is one of many 
defects in diamond and stands out due to its unique electronical and optical properties~\cite{Doherty2013,Bradac2019}. Optical pumping of the NV center 
yields photoluminescence in the visible spectrum and can initialize a specific spin 
state~\cite{Levine2019}. Using resonant microwave radiation, the spin state can be manipulated and then read out 
due to the spin-dependent photoluminescence~\cite{Levine2019}. \\
The NV center offers advantages like operation at room temperature and in ambient magnetic 
fields~\cite{Degen2017,Levine2019,Jelezko2006}. However, it is currently not possible to reach the necessary 
sensitivities for applications like magnetoencephalography ($<\SI{500}{\femto\tesla\per\sqrt{Hz}}$~\cite{Hämäläinen1993}), as 
low photoluminescence collection efficiency and 
scattering effects limit the signal strength and thus the sensitivity~\cite{Taylor2008,Rondin2014}. \\
To boost the sensitivity of NV magnetometers, Jeske et al.~\cite{Jeske2016} proposed the concept of laser 
threshold magnetometry (LTM). Here, an NV-doped diamond is used as an optically active laser medium relying on stimulated 
emission rather than spontaneous emission~\cite{Jeske2016}. The advantages of stimulated emission are high magnetic-field-dependent contrast 
and coherent signal strength yielding a magnetic-field-dependent laser threshold~\cite{Jeske2016,Dumeige2019,Webb2021,Gottesman2024}. \\
The first steps towards an experimental realization of LTM were the demonstration of stimulated emission from NV 
centers by Jeske et al.~\cite{Jeske2017} and increased magnetic field dependence due to stimulated emission as demonstrated by 
Hahl et al.~\cite{Hahl2022} inside a seeded cavity at a wavelength of \SI{710}{nm}. As it was not possible to demonstrate a continuous-wave laser cavity 
using only an NV-diamond as a gain medium~\cite{Hahl2022}, Lindner et al.~\cite{Lindner2024} realized a dual-media cavity containing an NV-doped 
diamond and a red semiconductor laser at a wavelength of \SI{690}{nm}. However, in this implementation, it was not possible to demonstrate a magnetic-field-dependent 
laser threshold~\cite{Lindner2024}. 
Instead of using the stimulated emission of the NV center Gottesman et al.~\cite{Gottesman2024} used the absorption of the singlet 
state at \SI{1042}{nm} and Schall et al.~\cite{Schall2025} showed similar absorption for shorter wavelengths. The absorption-based setups were 
not able to reach the predicted unity contrast~\cite{Gottesman2024}. \\
Here, we further investigate the approach of a dual-media cavity but using a different second laser medium. We 
realize a lasing system with a vertical external cavity surface emitting laser (VECSEL) emitting light at 
\SI{750}{nm}. The VECSEL medium has less gain and intrinsic losses compared to the previously used external cavity diode laser~\cite{Lindner2024}. This enables a cavity 
with higher finesse and more stability. The high finesse - combined with lower gain - should enhance the cavity output contribution of the NV centers compared 
to the VECSEL. VECSELs typically operate in a high finesse cavity with just a few percent output-coupling. This results in a long 
photon lifetime or long cavity decay time which can actually exceed the value of the upper-state lifetime. In this case, the VECSEL 
exhibits a class A laser dynamic~\cite{okhotnikov2010semiconductor}: 
The optical gain can react rapidly (in comparison to the dynamic of the NV centers) to variations in the photon density, leading to an exponential 
relaxation to the steady state and a strong damping and thus reduction of any intensity fluctuations~\cite{Jetter2021}. With this cavity, we demonstrate a 
magnetic-field-dependent laser threshold and support the 
corresponding measurements with an analytical description based on a rate model. This is the first realization of a magnetic-field-dependent 
threshold, which enables the development of a magnetometer based on LTM with a unity contrast. Furthermore, we 
point out that the commonly used formula for shot-noise-limited sensitivity in continuous-wave optically detected magnetic resonance (cw-ODMR) is an 
approximation for small magnetic field contrast and develop a general formula. This becomes relevant with the increasing 
contrast of LTM. In total, the results set the basis for NV-magnetometry with \SI{100}{\percent} contrast.
\section{\label{sec:results}Results}
\subsection{Two-media laser cavity}
\begin{figure}[h]
    \centering
    \includegraphics{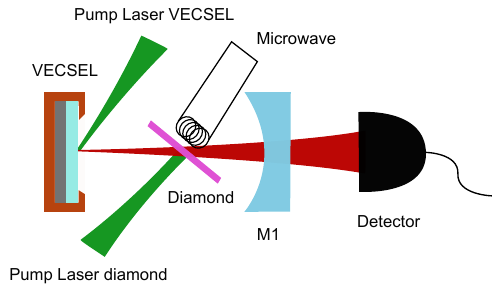}
    \caption[Setup]{Schematic of the two-media laser cavity. The optically active media are a vertical external cavity 
    surface emitting laser (VECSEL) and an NV-doped diamond. The cavity is formed by the back-side mirror of the VECSEL structure and a 
    concave mirror (M1). Both media are optically pumped with a focused green laser. The spin state of the NV centers is manipulated either with a permanent 
    magnet or a resonant microwave drive (Microwave). The cavity signal is detected with a powermeter or a photodiode connected 
    to a lock-in amplifier.}\label{fig:setup}
\end{figure}
Since previous efforts of building a two-media setup failed to produce a magnetic-field-dependence, it is important 
to choose the second medium carefully. The second active medium should provide sufficient gain to overcome the present cavity losses, but 
its gain should be on the same order of magnitude as the NV gain (cf. Eq.~\ref{eq:total cavity photons}). Otherwise, it would overshadow any effects of the NV 
centers. Moreover, we want to have a cavity with high finesse to further amplify the effect of the NV centers~\cite{Hahl2022,Schall2025}. Considering 
all of the above, we arrived at the experimental setup shown in Fig.~\ref{fig:setup}. The linear hemispherical optical 
cavity includes a VECSEL gain structure based on AlGaAs which emits red light ($\lambda_r = \SI{750}{nm}$) with an SiC heat spreader on the front. The cavity 
is formed between the back mirror inside the VECSEL structure and an external concave mirror (M1) with a radius of curvature $\mathrm{ROC} = \SI{50}{mm}$ 
with an outcoupling reflectivity $R=\SI{99.98}{\percent}$. 
The cavity is operated close to the stability limit with a total length of $l= \SI{50}{mm}$, which results in a cavity mode radius of $w_0 = \SI{50}{\micro m}$ on the 
VECSEL chip. Stability is assured as the diamond elongates the optical path. An NV-doped diamond with a thickness of $d = \SI{300}{\micro m}$ is placed $\SI{15}{mm}$ 
behind the VECSEL such that the cavity mode has a radius of $w_d = \SI{100}{\micro m}$ at the position of the diamond. The diamond plate is rotated to Brewster's angle 
to minimize reflection losses at the diamond surface and fixing the polarization. Both active media are pumped independently with green lasers ($\lambda_g = \SI{532}{nm}$). 
While the absorption cross section of the NV center is greater for longer wavelengths, $\lambda_g$ is selected due to its accessibility at low noise level. Cavity signals 
are detected behind the cavity mirror M1 either directly with a powermeter or using a photodiode connected to a lock-in amplifier. 
In order to manipulate the spin states, two methods are used. First, a permanent magnet is placed approximately \SI{1}{mm} 
above the diamond plate causing strong spin mixing analogous to applying a resonant microwave field. For 
frequency sweeps, a loop antenna is placed parallel to the diamond surface enclosing the cavity mode.
\subsection{Magnetic-field-dependent threshold\label{sec:thresh}}
\begin{figure}[t]
    \centering
    \includegraphics{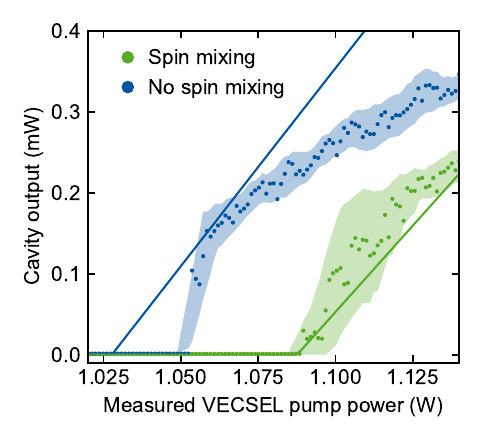}
    \caption[threshold]{Magnetic-field-dependent laser threshold. The cavity output is shown for a fixed diamond pump power of 
    $\SI{8}{W}$ with an illuminated area of $\SI{0.031}{mm^2}$ and varying VECSEL pump power with (green) and without (blue) 
    spin mixing using a permanent magnet. The dots represent the measurement values and the shadows represent the standard 
    deviation from five measurements. The lines show an simulation with parameters based on the experimental conditions. 
    The cavity loss parameter was used as the only fit parameter to match the lines to the experimental data.}\label{fig:threshold}
\end{figure}
LTM enhances the signal change from a magnetic field by using the competition between 
spontaneous and stimulated emission, given that the stimulated emission rate changes with the intensity of the 
field in the cavity. Or in other words the optical non-linearity of a laser cavity is used to enhance the sensor signal. 
This nonlinearity is strongest when close to the threshold~\cite{Jeske2016,Lindner2024,Gottesman2024}. \\
To benefit from this, we aim at a laser operating regime where the laser threshold is shifted by a magnetic field. If we then use a pump power between 
these two thresholds, \SI{100}{\percent} contrast can be 
realized and thus the strongest signal enhancement is possible. This is the foundation for a strong improvement in 
sensitivity. A two-media cavity can provide the additional gain from a second laser medium which is necessary to 
achieve lasing as the gain of the NV centers is not sufficiently high to compensate the cavity losses. Although the diamond provides optical gain, it 
also has intrinsic and pump induced optical losses, which make a second active medium necessary~\cite{Hahl2022}. LTM can then be achieved by 
magnetic-field-dependent gain from the NV-diamond. Thus we aim to achieve a laser system with a well defined 
threshold and then check for the magnetic-field dependence of the threshold. \\
Since there are two gain media in the cavity, a laser threshold can be achieved by either fixing the pump power on 
the VECSEL and varying the pump power on the diamond or vice versa. Although only the NV diamond's gain is magnetic-field-dependent, the change in gain can be expected 
to shift the laser threshold for both configurations. There even is an advantage in fixing the diamond pump power and measuring the laser threshold as a function of 
the pump power of the VECSEL: we can choose a large diamond pump power which ensures strong spin polarization of the diamond when no spin mixing is 
applied~\cite{Scheuer2016} and thus maximize the magnetic contrast due to spin mixing. Also taking into account that the VECSEL is a stronger gain 
medium and thus a threshold is reached for smaller VECESEL pump powers, leads to the conclusion to fix the diamond pump power at \SI{8}{W} and vary the VECSEL pump power. \\
We then investigate the magnetic-field-dependence by measuring with and without a strong magnetic field 
present using a permanent magnet. The field component perpendicular to the NV axis leads to a mixing of 
the spin states and thus the measurement with and without the magnet is a measurement with NVs in the dark spin 
state or in the bright $m_S = 0$ spin state, respectively. Applying a magnetic field with strong perpendicular components 
is similar to applying resonant microwaves and yields higher contrast due to greater field strength and less inhomogeneity~\cite{HahlPHD}. \\
The cavity output as a function of the VECSEL pump power with and without spin mixing via a permanent magnet is shown in 
Fig.~\ref{fig:threshold}. Measurements are represented by dots and the shadow represents the standard deviation from five 
measurements. We can distinctly observe a shift of the lasing threshold when spin mixing is introduced. This marks the 
first measurement of a laser threshold shifted by a magnetic field due to decreased NV center gain. Both curves show an initially steep slope above threshold 
that decreases with higher VECSEL pump power. This can be understood in terms of the interplay of spontaneous and stimulated emission and leads to a stronger 
increase of cavity photons directly at the threshold (cf. \textcite{Milonni2010}, p. 220ff). 
Moreover, we simulated the system based on a rate model (cf. method section~\ref{sec:rate model}) using parameters based on the experimental 
conditions. Only the cavity loss rate was used as a fit parameter to yield the best overlap with the experimental results. 
The optimized value is $\kappa_{sim} = \SI{49.5}{MHz}$, which is the same order of magnitude as our estimated values of 
$\kappa_{est} = \SI{31}{MHz}$. Simulation results are represented by lines in Fig.~\ref{fig:threshold} and show a good 
agreement with the experimental data. Differences in the slope of the cavity output and the need for optimization of the cavity loss rate
indicate an additional pump-laser-induced absorptive effect, which will be discussed in the following section.  
\subsection{\label{sec:ind-ab}Induced absorption}
\begin{figure}[h]
    \centering
    \includegraphics{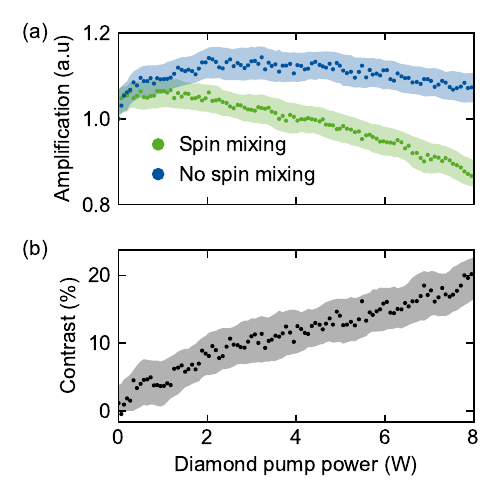}
    \caption[induced absorption]{Induced absorption and magnetic-field-dependence.~(a) Amplification of an initial 
    cavity output is shown for a constant VECSEL pump power and varying diamond pump power. The measurement was done 
    with (green) and without (blue) spin mixing by a permanent magnet.~(b) Relative contrast between the curves from (a). 
    }\label{fig:ind ab}
\end{figure}
A possible effect causing the difference between simulations and measurement results may be induced absorption. 
This effect has been initially observed by \textcite{Hahl2022} in a seeded cavity setup at a comparable 
wavelength of \SI{710}{nm}. Its exact physical origin and behaviour still remains unclear. When pumping 
the diamond, an absorption occurs that counteracts the stimulated emission from the NV centers and even leads 
to a reduction of the cavity output for high diamond pump powers~\cite{Hahl2022}. To validate our assumption, we fix the VECSEL 
pump power at a constant value such that the cavity has a stable but low output of \SI{1}{mW} when the NV centers are not being pumped. Then we 
measure changes in cavity output when varying the diamond pump power. \\
The result is shown in Fig.~\ref{fig:ind ab}~a. We can see an amplification of the initial 
signal of $\SI{16}{\percent}$ by stimulated emission from the NV centers, followed by a decrease to only $\SI{7}{\percent}$ amplification for pump powers higher 
than $\SI{4}{W}$. This behavior 
matches the observations made by Hahl et al.~\cite{Hahl2022}, such that we can attribute this behavior to induced 
absorption. Placing a permanent magnet close to the diamond sample results in a spin mixing similar to a resonant 
microwave drive due the transverse component of the field with respect to the axis of the NV center. Repeating the 
measurement now including such a spin mixing with a permanent magnet leads to stronger 
absorption than in the measurement without spin mixing. This reduces the cavity signal by up to $\SI{13}{\percent}$. 
The contrast between the measurements with and without spin mixing, which is depicted in Fig.~\ref{fig:ind ab}~b, 
maximizes for increasing diamond pump power. Despite the high pump power the intensity is limited due to 
the large illuminated area of $\SI{0.031}{mm^2}$. Thus for a pump power of e.g. $P_{\mathrm{NV}} = \SI{5}{W}$ we 
expect an excitation rate of $\Lambda_{\mathrm{NV}} = \SI{5.75}{MHz}$ which only leads to a partial spin polarisation 
of the spin triplet state of \SI{56.88}{\percent} according to our rate model outlined below. Thus we attribute the increasing contrast 
with diamond pump power to higher spin polarization without spin mixing, which leads to larger changes in the cavity 
output power. In the seeded cavity, the contrast reached a maximum at a certain diamond pump power, which we do not 
observe here, as the pump intensities are not high enough~\cite{Hahl2022}.
\subsection{\label{sec:odmr}Microwave-based measurements}
After investigating the behavior due to magnetic fields, we want to evaluate the performance of the setup as a magnetometer. In the following experiments, 
a microwave drive with a signal strength of \SI{40}{dBm} is used for the spin manipulation and ODMR is performed. For this, the cavity output is detected 
while sweeping the frequency of the applied microwave radiation. In comparison to using a permanent magnet, the spin manipulation by microwave radiation provides 
lower spin mixing and therefore lower contrast due to limited power of the generated signal and field inhomogeneity. Additionally, the microwave drive is modulated in 
amplitude and the signal is analyzed with a lock-in amplifier, yielding a higher signal-to-noise ratio. The resulting lineshape is a Lorentzian and 
its parameters directly determine the theoretical limit of the shot-noise-limited sensitivity. The corresponding formula, which is predominant in the field, was proposed 
by \textcite{Dreau2011}(Eq.~3) and calculates the shot-noise-limited sensitivity as
\begin{equation}
    \eta_c = \frac{2 \pi}{\gamma_e} \frac{8}{3\sqrt{3}} \frac{\delta}{C \sqrt{I_0}}.
\end{equation}
Here, $\delta$ is the half width at half maximum, $C$ the contrast, $I_0$ the signal baseline and $\gamma_e$ 
the gyromagnetic ratio. However, its derivation includes the assumption that the signal strength at the operating 
point is approximately equal to the signal baseline. This assumption is only valid for small contrast. Therefore, we 
include a scaling factor $S_F$ to arrive at a generalized formula:

\begin{equation}
    \eta_{shot} = \eta_c \underbrace{\frac{\sqrt{(3+S_C^2)^3 (3+S_C^2 - 3C)}}{16 S_C}}_{:= S_F},
\end{equation}
with
\begin{equation}
    S_C = \sqrt{C -1 + \sqrt{C^2 - 5C +4}},
\end{equation}
which is derived in the method section~\ref{sec:sens}. It is important to note that the factors $S_F$ and $S_C$ depend on the measurement method. The results 
above are given for cw-ODMR, however the factors for amplitude-modulated (AM) ODMR are also derived in section~\ref{sec:sens}. The scaling factor $S_F$ solely 
depends on the contrast $C$ and improves the sensitivity with increasing contrast. As the formula by \textcite{Dreau2011} scales linearly with contrast, the correction 
leads to an overall superlinear scaling of the sensitivity with regard to the ODMR contrast. These results indicate that the sensitivity improvements of LTM, which 
rely on high contrasts, may be even higher than expected. \\
\begin{figure}[t]
    \centering
    \includegraphics{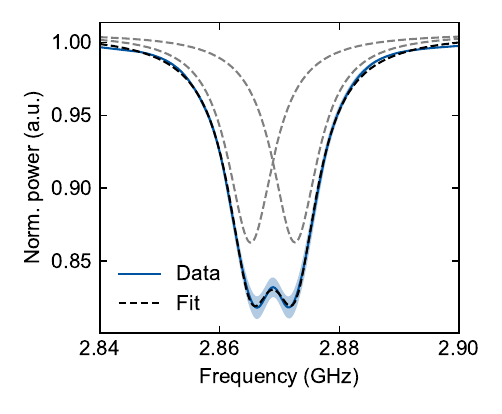}
    \caption[ODMR]{Optically detected magnetic resonance. Normalized cavity output is shown over the frequency 
    of the microwave drive. The data (blue) is fitted with a double Lorentzian (black), for which the single Lorentzians 
    are also shown (gray dashed).}\label{fig:ODMR}
\end{figure}
Figure~\ref{fig:ODMR} shows the resulting ODMR spectrum normalized to the signal baseline of $\SI{650}{\mu W}$. On resonance, a total contrast of $\SI{19}{\percent}$ is 
reached. As it was possible to shift the lasing threshold with the permanent magnet, we would expect to reach a unity contrast of unity. However, we could not achieve 
such a high contrast in the ODMR experiment. This becomes clear when comparing the signal baseline of $\SI{650}{\mu W}$ with the output powers observed in 
Fig.~\ref{fig:threshold}. The output power in the region of unity contrast is considerably lower and was not stable enough for performing ODMR. Furthermore, the loop 
antenna has some field inhomogeneity along the comparatively large thickness of the diamond sample ($\SI{300}{\mu m}$), which also reduces the threshold shift. Therefore, 
it is reasonable to assume that in an improved setup, it will be possible to observe higher contrast. We fitted a double Lorentzian function to the data which is shown in 
black. Based on the parameters of a single Lorentzian and using Eq.~\ref{eq:sens_am}, a shot-noise-limited sensitivity of $\SI{49.07 \pm 0.33}{\pico\tesla\per\sqrt{Hz}}$ is 
calculated. For this calculation it is assumed that a magnetic field with [100] orientation is applied to the NV-diamond sample, such that all NV orientations contribute 
to the contrast (cf. \textcite{Barry2023} and \textcite{Graham.2023}).
\section{Conclusions}
This work successfully demonstrates the measurement of a magnetic-field-dependent threshold in a dual-media cavity setup 
incorporating an NV-doped diamond and a VECSEL. This magnetic-field dependent threshold is an essential prerequisite 
and sets the basis to achieve up to $\SI{100}{\percent}$ ODMR contrast in future laser threshold magnetometers. The results 
show a significant correlation between the experimental data and simulations regarding the magnetic field dependence. 
However, it is important to consider the effects of induced absorption observed in this configuration. In the future, 
investigating this effect further may yield a full description of the physical processes inside the NV center. We 
have derived a generalized formula for calculating shot-noise-limited sensitivity in ODMR, which was applied to 
evaluate the performance of our setup, yielding a shot-noise-limited sensitivity of $\SI{49.07 \pm 0.33}{\pico\tesla\per\sqrt{Hz}}$. 
While the sensitivity is currently constrained, the use of an optimized diamond with significantly higher NV-concentration 
suggests sensitivities down to \SI{4.9}{\femto\tesla\per\sqrt{Hz}} (cf. Sec.~\ref{sec:rate model}), which would greatly improve 
current solutions. Future work focusing on refining the microwave implementation like diamond coatings and novel microwave 
resonators are expected to enhance the contrast and overall performance of the system.
\section{Methods}
\subsection{Rate model}\label{sec:rate model}
\begin{figure}[h]
    \centering
    \includegraphics{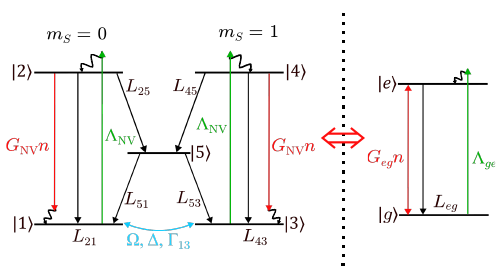}
    \caption[Rate model]{Level scheme assumed in the rate model describing the two-media cavity. The NV center is modelled 
    as five states with excitation (green), spontaneous emission (black), stimulated emission (red), and coherent 
    microwave drive (blue). Squiggly lines represent phononic transitions. The VECSEL is modelled as a two-level 
    emitter with off-resonant pumping.}\label{fig:rate model}
\end{figure}
The theoretical model to analytically simulate the two-media system is derived 
by calculating the equations of motion (EOM) for the density matrices of the NV 
center and the VECSEL. Also taking into account the dynamics of the cavity photons 
yields a description of the cavity output in terms of fundamental NV parameters. The 
EOM are calculated using the Lindblad master equation. All NV and VECSEL states are depicted 
in Fig.~\ref{fig:rate model}. States $\ket{1}$ and $\ket{3}$ represent the $^3$A$_2$ triplet 
ground states. The microwave drive between them is described as a Rabi oscillation with 
strength $\Omega$, detuning $\Delta$, and dephasing $\Gamma_{13}$. Exciting the electrons from 
the ground state with a green laser is shown as the green arrows with rate $\Lambda_{\mathrm{NV}}$. 
As the excitation energy is larger than the zero phonon line (ZPL), the electrons are excited in 
a phonon sideband from where they relax into the $^3$E triplet states, which are here states 
$\ket{2}$ and $\ket{4}$. The phononic decay is depicted using squiggly lines since the process is 
much faster than the other processes and therefore not included in the rates~\cite{Doherty2013}. 
The NV centers can relax either via the singlet state that is denoted by $\ket{5}$ or directly back into 
the triplet ground states. 
When decaying directly, we have to differentiate between the broadband spontaneous emission, that hardly 
emits into the cavity mode (black arrows) and the stimulated emission at \SI{750}{nm} (red arrows). 
As the stimulated emission has less energy than the ZPL, the electron decays in a phonon sideband, which 
means that an absorption of cavity photons is highly unlikely and therefore not considered in the model. 
The stimulated emission rate is defined by the cavity coupling $G_{\mathrm{NV}}$ multiplied with the number 
of cavity photons (per NV centre) n. The cavity coupling is calculated with~\cite{siegman86}
\begin{equation}
    G_{\mathrm{NV}} = \frac{3 \nu \Gamma \lambda^3 N_{\mathrm{NV}}}{4 \pi^2 \Delta \nu n_r^3 V_c} ,
\end{equation}
with transition frequency $\nu$, spontaneous transition rate $\Gamma$, wavelength $\lambda$, number of NV 
centers $N_{\mathrm{NV}}$, transition width $\Delta\nu$, refractive index of diamond $n_r$, and cavity volume $V_c$.
In order to use the same formalism for both media, we model the VECSEL for simplicity as a collection of two-level 
emitters which are pumped off-resonantly. However, this model should describe the medium accurately in the low-power 
regime, where thermal effects are negligible. To estimate the number of ``VECSEL emitters'', we use 
typical electron surface densities and the geometric structure of the experimental setup~\cite{Koch2014}. 
The other VECSEL parameters are calculated with cavity loss and threshold pump power for the cavity with 
and without the diamond. Afterwards, the cavity loss rate is optimized to yield good overlap with the experimental 
data. All parameters used for the simulation are shown in Tab.~\ref{tab:params}.
To obtain the analytical solution the same procedure as is used in Jeske et al.~\cite{Jeske2016}, with 
the difference that the equation for the cavity photons contains now contributions from both media. 
Computing the steady state allows to solve the master equation for the two media independently as a function of 
the cavity photons. Inserting the solutions into the differential equation for the cavity photons yields 
a single equation for the steady state number of cavity photons. The set of equations of motion, with $N$ 
being the total number of cavity photons reads 
\begin{widetext}
\begin{subequations}
    \begin{align}
        \droh{13} & = \left( \ima \Delta - \Gamma_{13} - \lnv \right) \rho_{13} + \ima \Omega \left( \rho_{11} - \rho_{33} \right), \\
        \droh{31} & = - \left( \ima \Delta + \Gamma_{13} + \lnv \right) \rho_{31} +  \ima \Omega \left( \rho_{33} - \rho_{11} \right), \\
        \droh{11} & = \ima \Omega \left( \rho_{13} - \rho_{31} \right) - \lnv \rho_{11} +  \left( \lij{21} + \gnv N / N_{\mathrm{NV}} \right)  \rho_{22} + \lij{51} \rho_{55}, \\
        \droh{22} & = \lnv \rho_{11} - \left( \lij{21} + \gnv N / N_{\mathrm{NV}} + \lij{25} \right) \rho_{22}, \\
        \droh{33} & = \ima \Omega \left( \rho_{31} - \rho_{13} \right) - \lnv \rho_{33} + \left( \lij{43} + \gnv N / N_{\mathrm{NV}} \right)  \rho_{44} + \lij{53} \rho_{55}, \\
        \droh{44} & = \lnv \rho_{33} - \left( \lij{43} + \gnv N / N_{\mathrm{NV}} + \lij{45} \right) \rho_{44}, \\
        \droh{55} & = \lij{25} \rho_{22} + \lij{45} \rho_{44} - \left( \lij{51} + \lij{53} \right) \rho_{55}, \\
        \droh{gg} & = - \left( \Lambda_{ge} + G_{eg} N / N_{\mathrm{V}} \right) \rho_{gg} + \left(L_{eg} + G_{eg} N / N_{\mathrm{V}} \right) \rho_{ee}, \\
        \droh{ee} & = \left( \Lambda_{ge} + G_{eg} N / N_{\mathrm{V}} \right) \rho_{gg} - \left(L_{eg} + G_{eg} N / N_{\mathrm{V}} \right) \rho_{ee}, \\
        \dot{N}   & = \gnv N \left( \rho_{22} + \rho_{44} \right) + G_{eg} N \left( \rho_{ee} - \rho_{gg} \right) - \kappa N. \label{eq:total cavity photons}
    \end{align}
\end{subequations}
\end{widetext}

The formalism not only enables us to simulate the optical cavity output, but also the sensitivity using Eq.~\ref{eq:sens-gen}. When aiming 
to optimize the sensitivity, the relevant parameters are the diamond pump power $P_{\mathrm{NV}}$, the VECSEL pump power $P_{\mathrm{V}}$, 
and the applied Rabi frequency $\Omega$. Given a pair of $P_{\mathrm{NV}}$ and $\Omega$, we have to choose the right VECSEL pump power, such 
that we operate the system close to the laser threshold. Avoiding numerical complications when approaching unity contrast, we choose the VECSEL 
pump power such that the contrast is $C= \num{0.99}$. This yields an optimal shot-noise-limited sensitivity of \SI{5.13}{\pico\tesla\per\sqrt{Hz}}. 
To explore the limitations of two-media LTM, we repeated the optimization for a diamond with a significantly higher NV-concentration of \SI{16}{ppm} 
and $T_2^\star$ of \SI{181}{ns} as already used in Refs.~\cite{Acosta2009,Jeske2016,Lindner2024}. Due to the higher NV concentration, a shot-noise-limited 
sensitivity of \SI{4.9}{\femto\tesla\per\sqrt{Hz}} was reached. 
\begin{table}
\caption[Simulation parameters]{Parameters used for the simulation with the rate model}
\begin{ruledtabular}
    \begin{tabular}{ccc}
        Parameter & Value & Reference \\
        \hline
        \rule{0pt}{1\normalbaselineskip}$L_{21}$ & \SI{66.16}{MHz} & \cite{Gupta2016} \\
        $L_{43}$ & \SI{66.16}{MHz} & \cite{Gupta2016} \\
        $L_{25}$ & \SI{11.1}{MHz} & \cite{Gupta2016} \\
        $L_{45}$ & \SI{91.8}{MHz} & \cite{Gupta2016} \\
        $L_{51}$ & \SI{4.87}{MHz} & \cite{Gupta2016} \\
        $L_{53}$ & \SI{2.04}{MHz} & \cite{Gupta2016} \\
        $\Omega$ & \SI{3.5}{MHz} \\
        $\Gamma_{13}$ & \SI{1}{MHz} \\
        $\Lambda_{\mathrm{NV}} / P_{\mathrm{NV}}$ & \SI{1.15}{MHz \per W} & \cite{Subedi2019} \\
        $G_{\mathrm{NV}}$ & \SI{230}{MHz}  \\
        $N_{\mathrm{NV}}$ & \num{1.27e12} \\ 
        $L_{eg}$ & \SI{30}{MHz}  \\
        $\Lambda_{ge} / P_{\mathrm{V}}$ & \SI{47}{MHz \per W} \\ 
        $G_{eg}$ & \SI{130}{MHz}  \\
        $N_{\mathrm{V}}$ & \num{5.3e10} & \cite{Koch2014}\\
        $\kappa_{est}$ & \SI{31}{MHz} \\
        $\kappa_{opt}$ & \SI{49.5}{MHz} \\
    \end{tabular}
    \label{tab:params}
\end{ruledtabular}
\end{table}
\subsection{Components and devices}
The diamond used in this setup is a 3x3x0.3 mm commercial high pressure high temperature (HPHT) [100]-oriented diamond with low pressure 
high temperature (LPHT) treatment, irradiation, and annealing. The NV concentration can be estimated to \SI{1.86}{ppm} and the sample has an absorption 
of about \SI{0.01}{\per\centi\meter}, see Ref.~\cite{Hahl2022,Schall2025}. It is pumped with a Spectra-physics 
MillenniaXsS. The VECSEL gain structure is based on AlGaAs/GaAs material and is pumped with a Novanta axiom532. The mirror 
at the end of the cavity is a Layertec 100974 dielectric mirror.
 
In Sections~\ref{sec:thresh} and \ref{sec:ind-ab}, a permanent magnet (Neodymium, $\SI{10}{mm}$ diameter, $\SI{3}{mm}$ height) is used for providing a strong 
magnetic field for spin mixing. To manipulate the NV spin states in Section~\ref{sec:odmr}, a single loop antenna with a diameter of about \SI{1}{mm} was used. 
The signal was generated using a Rohde \& Schwarz SMBV100B with a power of \SI{-5}{dBm} and amplified by \SI{45}{dB} with a Mini-Circuits ZHL-16W-43-S+ to a total 
power of \SI{40}{dBm}. This high signal strength is necessary as the loop antenna cannot be impedance matched. To detect the cavity output either a powermeter 
(Thorlabs S130C) with the longpass filters Thorlabs FELH0700 and Thorlabs FELH0600, or a OE-300-SI-10-FST photodiode with 
filters Semrock BLP01-532R-25 and Semrock NF533-17 were used. The photodiode is connected to the lock-in amplifier Zurich Instruments MFLI 500 kHz.
\subsection{Shot-noise-limited sensitivity for cw ODMR}\label{sec:sens}
\begin{figure}[htp]
	\centering
	\includegraphics{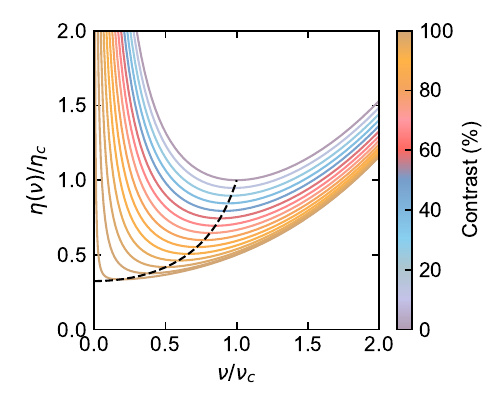}
	\caption[around voof]{Normalized frequency dependent sensitivity around the inflection point 
			for different contrasts. The sensitivity is normalized by $\eta_c$ (Eq.~\ref{eq:sens-com}) and 
			the frequency is given in units of the inflection point frequency $\nu_c$. The black dotted line shows the optimal operating frequency and corresponding normalized 
			sensitivity.}\label{fig:aroundvoof}
\end{figure}

\begin{figure}[htp]
   \centering
   \includegraphics{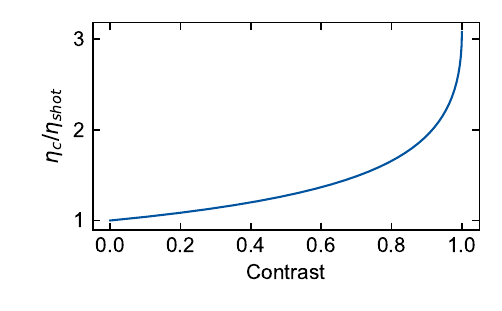}
   \caption[Proportional]{Ratio of sensitivity formula by \textcite{Dreau2011} ($\eta_c$) and newly derived shot-noise-limited sensitivity 
   at the OOF ($\eta_{shot}$). The ratio is solely dependent on the contrast.}\label{fig:proportional}
\end{figure}

The sensitivity of a DC magnetometer is given by~\cite{Taylor2008}
\begin{equation}
    \eta_{DC} = \sigma_B \sqrt{T},
\end{equation}
with magnetic field standard deviation $\sigma_B$ and measurement time $T$. As we measure 
photons with a detector, $\sigma_B$ can be rewritten using error propagation as 
$\sigma_B = \abs{\partial B/ \partial I} \sigma_I$, where $I$ is the measurement signal in photons per second. 
Counting photons follows the Poisson statistic which means that we can rewrite the 
error on our measurement signal in the photon shot-noise limit as 
$\sigma_I \sqrt{T} = \sqrt{I}$. At this point, \textcite{Dreau2011} make 
the assumption that the error can be fixed to the baseline power $I_0$ of a cw-ODMR spectrum, 
which is defined as 
\begin{equation}
    I(\nu) = I_0 \left( 1 - \frac{C \delta^2}{\delta^2 + \nu^2} \right).
\end{equation}
There, $\nu$ is the frequency shift from the resonance, $C$ the contrast, and $\delta$ the half 
width at half maximum. Using this, they obtain the shot-noise-limited sensitivity (\textcite{Dreau2011}, Eq.~3)
\begin{equation}\label{eq:sens-com}
    \eta_c = \frac{2 \pi}{\gamma_e} \frac{8}{3\sqrt{3}} \frac{\delta}{C \sqrt{I_0}}
\end{equation}
by choosing the operating frequency of the magnetometer at the inflection point $\nu_c = \delta / \sqrt{3}$ of 
the cw-ODMR spectrum~\cite{Vanier1989}. The signal strength at this inflection point is $I(\nu_c) = I_0 (1-3C /4)$. 
Therefore, Eq.~\ref{eq:sens-com} is only valid as long as $I(\nu_c) \approx I_0$ and the point of optimal sensitivity 
in the curve is approximately at the inflection point, which are true in the regime of low contrast. 
For setups with high contrast, as for LTM, we have to calculate the point of optimal sensitivity on the curve 
and determine the sensitivity at this point without approximations. The frequency-dependent sensitivity is then given by 
\begin{equation}\label{eq:sens-gen}
    \eta (\nu) = \sqrt{I(\nu)} \abs{\frac{\partial B}{\partial I(\nu)}} = \frac{2 \pi}{\gamma_e} \sqrt{I(\nu)} \abs{\frac{\partial I(\nu)}{\partial \nu}}^{-1},
\end{equation}
where the Zeeman splitting gives the proportionality between magnetic field and frequency shift 
in form of the gyromagnetic ratio $\gamma_e$. A consequence of the additional frequency-dependence is a different 
optimal operating frequency (OOF). The new  OOF can be determined by taking the derivative of Eq.~\ref{eq:sens-gen} 
and calculating its roots
\begin{equation}
    \nu_{OOF} = \nu_c  \underbrace{\sqrt{C - 1 + \sqrt{C^2 - 5C + 4}}}_{=: S_C}.
\end{equation}
Inserting $\nu_{OOF}$ into Eq.~\ref{eq:sens-gen}, then yields the shot-noise-limited sensitivity
\begin{equation}\label{eq:sens-shot}
    \eta_{shot} = \eta (\nu_{OOF}) = \eta_c \frac{\sqrt{(3+S_C^2)^3 (3+S_C^2 - 3C)}}{16 S_C} .
\end{equation}
We can see that the general formula scales Eq.~\ref{eq:sens-com} with a factor that is solely 
dependent on the contrast. Both, the shift of the OOF as well as the improved sensitivity can be 
observed in Fig.~\ref{fig:aroundvoof}. Here, the frequency dependent sensitivity is normalized by the 
commonly used formula and shown around the inflection point at $\nu = \nu_c$. The black dotted line shows 
the OOF and corresponding shot-noise-limited sensitivity for the respective contrasts. In the limit of 
unity contrast, the OOF approaches the resonance, and the normalized sensitivity goes to $3\sqrt{3}/16$. 
The inverse normalized shot-noise-limited sensitivity at the OOF is shown in Fig.~\ref{fig:proportional}. 
There, the improvement of the sensitivity for increasing contrast becomes clear. As $\eta_c$ already scales 
linearly with the contrast using the generalized formula, this result shows that the correct formula for the 
shot-noise-limited sensitivity $\eta_{shot}$ scales superlinearly with the contrast. Thus an increase in contrast 
from \SI{10}{\percent} to \SI{100}{\percent} will give an improvement in sensitivity by a factor of more than \num{10}. 
The improvement factor for this example is \num{29.6}.

So far, this discussion only includes considerations for cw-ODMR. In order to give an estimation of the 
shot-noise-limited sensitivity, we have to account for the measurement method. For AM-ODMR, 
the signal is given as $I_{AM}(\nu) = \chi (I_0 - I(\nu))$, with some proportionality constant $\chi$. By replacing 
$I(\nu)$ with $I_{AM}(\nu)$ in Eq.~\ref{eq:sens-gen} and following the same procedure as above we arrive at two 
new scaling factors for the OOF and the shot-noise-limited sensitivity. The scaling factor for the OOF is
\begin{equation}
    S_{C,AM} = \sqrt{\frac{C - 2 + \sqrt{C^2 -10C + 16}}{2}}
\end{equation}
and the shot-noise-limited sensitivity for AM-ODMR reads
\begin{align}\label{eq:sens_am}
    \eta_{shot,AM} &= \eta_c \, S_{F,AM} \\ &= \eta_c \frac{\sqrt{(3+S_{C,AM}^2)^3 (6+2 S_{C,AM}^2 - 3C)}}{16 S_{C,AM}}.
\end{align}
For amplitude modulation, the simulations show a similar improvement of the shot-noise-limited sensitivity in comparison to 
cw-ODMR, however the exact dependency on the contrast depends on the read-out and measurement technique.
\section*{Acknowledgments}
We acknowledge funding from the Bundesministerium für Bildung und Forschung (13N16485), 
Fraunhofer-Gesellschaft zur Förderung der angewandten Forschung e.V. (QMag, QMag - Next Level), 
and the Ministerium für Wirtschaft, Arbeit und Wohnungsbau Baden-Württemberg (QMag). The authors 
would like to thank Steffen Adler and Peter Holl for valuable discussions regarding the experimental
setup.
\section*{Author Declarations}
\subsection*{Conflict of Interest}
F.A.H., J.J., and T.L. have a patent related to the content filed by Fraunhofer-Gesellschaft zur 
Förderung der angewandten Forschung e.V. (DE102021209666A1) pending.\\
J.J. and M.R. have a patent related to the content filed by Fraunhofer-Gesellschaft zur 
Förderung der angewandten Forschung e.V. (DE102020204022A1) pending.
\subsection*{Author Contributions}
Y.R. and L.L. contributed equally to this work.
\\ 
\textbf{Yves Rottstaedt}: Conceptualization (equal); Data curation (equal); Formal analysis (equal); Methodology (equal); Software (equal); Validation (equal); Visualization (equal); Writing – original draft (lead); Writing – review \& editing (equal).
\textbf{Lukas Lindner}: Conceptualization (equal); Data curation (equal); Formal analysis (equal); Methodology (equal); Software (equal); Validation (equal); Visualization (equal); Writing – original draft (supporting); Writing – review \& editing (equal).
\textbf{Florian Schall}: Methodology (equal); Validation (equal); Writing – review \& editing (equal).
\textbf{Felix A. Hahl}: Software (equal); Validation (equal); Writing – review \& editing (equal).
\textbf{Tingpeng Luo}: Resources (equal); Writing – review \& editing (equal).
\textbf{Florentin Reiter}: Validation (equal); Writing – review \& editing (equal).
\textbf{Takeshi Ohshima}: Resources (equal); Writing – review \& editing (equal).
\textbf{Alexander M. Zaitsev}: Resources (equal); Writing – review \& editing (equal).
\textbf{Roman Bek}: Resources (equal); Writing – review \& editing (equal).
\textbf{Marcel Rattunde}: Conceptualization (equal); Methodology (equal); Supervision (equal); Funding acquisition (equal); Writing – review \& editing (equal).
\textbf{Jan Jeske}: Conceptualization (equal); Data curation (equal); Methodology (equal); Validation (equal); Project administration (equal); Supervision (equal); Funding acquisition (equal); Writing – review \& editing (equal).
\textbf{Rüdiger Quay}: Conceptualization (equal); Supervision (equal); Funding acquisition (equal); Writing – review \& editing (equal).
\subsection*{Data Availability}
The data that support the findings of this study are available from the corresponding author 
upon reasonable request.
%

%


%
%

%


\bibliography{refs}

\end{document}